# Quantum Photonics Incorporating Color Centers in Silicon Carbide and Diamond

Marina Radulaski [1], Jelena Vučković [1,*]

[1] E. L. Ginzton Laboratory, Stanford University, Stanford, CA 94305, USA

* E-mail: jela@stanford.edu

**Abstract**: Quantum photonics plays a crucial role in the development of novel communication and sensing technologies. Color centers hosted in silicon carbide and diamond offer single photon emission and long coherence spins that can be scalably implemented in quantum networks. We develop systems that integrate these color centers with photonic devices that modify their emission properties through electromagnetically tailored light and matter interaction.

## 1. Introduction

As the growth of personal and super- computing is nearing the limits of the so-called Moore's law [1], the paradigm where electronic information carriers are replaced with photons opens up new frontiers. Light-based computing holds the promise of achieving higher clock speeds through faster modulation of signal [2], reduced power consumption via harnessing of the strong light-matter interaction [3], exploration of novel architectures based on light interference effects such as the neuromorphic computing [4] and Ising solvers [5], and developing quantum computation and quantum cryptography platforms based on electron and nuclear spin-qubits [6] that would reduce algorithm complexity [7] and increase the safety of communication channels [8]. In parallel, these technologies allow for *in vivo* sensing of local index of refraction [9] and temperature [10], providing unprecedented nanoscale resolution measurements.

In order to engineer optical quantum systems and devices, one first needs to develop reliable photonic elements: waveguides to carry the signal [11], low-loss microresonators to enhance the local electromagnetic field intensity [12] and serve as nonlinear elements for switching [3] and frequency conversion [13], and narrow linewidth quantum emitters to serve as single photon sources or qubits [14]. Many of these components have been demonstrated in atomic optical systems harnessing cavity quantum electrodynamic effects, often by coupling rubidium atoms to a high finesse cavity [15]. However, the coupling rates in atomic systems are in the MHz range, the operation requires cryogenic temperatures and high vacuum. A leap forward has been achieved with implementations in III-V solid-state platforms incorporating quantum wells and quantum dots coupled to Distributed Bragg Reflector (DBR) and Photonic Crystal (PC) cavities [16]. These systems reach up to 25 GHz operating speeds and have on-chip footprint in tens of micrometers.

More recently, color centers in diamond and silicon carbide have taken a prominent position among solid-state quantum emitters, due to their narrow inhomogeneous (ensemble) linewidth, up to room temperature operation, and lattice-point size [14]. These properties make them ideal candidates for multi-emitter cavity quantum electrodynamics, a regime that has not yet been thoroughly explored in solid-state systems and could potentially increase the strength of interaction – and therefore the speed of optical devices – by an order of magnitude [17]. Additionally, color centers provide the optical readout of their electron spin, as well as the microwave spin-control with coherence times up to several seconds [18]. This property has been explored for the quantum bit (qubit) applications, as well as for the nanoscale magnetic and temperature measurements.

Color centers incorporated into photonic devices such as microcavities or waveguiding elements experience modified density of states that the photons are emitted to [12]. Depending on the strength of interaction, such coupling can result in faster and spectrally better defined single photon emission, or radically change the emission spectrum and statistics. Experimental realization of such devices requires the development of advanced growth methods and nanofabrication techniques.

We introduce approaches and devices in silicon carbide and hybrid silicon carbide-nanodiamond platforms that modify the emission of color centers for applications in optical networks and high sensitivity nanoscale sensing.

## 2. Photonic Substrate Properties

Photonic substrates are expected to provide mechanisms for light confinement, light generation and frequency conversion. Light confinement into a 2D slab is often facilitated by the total internal reflection (TIR), which relies on index of refraction contrast between the substrate and its environment. For example, photonics in silicon-on-insulator (SOI) is based on the TIR between silicon, $n_{Si}$ = 3.5, silicon dioxide, $n_{SiO2}$ = 1.5, from its bottom side, and air, $n_{air}$ = 1, from the top. To be able to confine light with low absorption, photon energy should be below the substrate band gap. Consequently, wider band gap materials offer a multitude of operational frequencies, including visible wavelengths, which are of great interest for biological applications due to the overlap with the water optical transparency window at the wavelength range 300-700 nm.

Photonics applications such as quantum cryptography – which relies on single photon operation [8]; and *in vivo* bio labeling – which requires bright point sources of light [19], can be implemented using solid-state quantum emitters. Quantum dots and color centers have been studied for these purposes. For example, 20 nm x 20 nm x 5 nm InAs quantum dots grown inside GaAs slabs confine electrons and holes inside the band gap that is narrower than the band gap of the surrounding substrate [20]; the electrons and holes recombine to valence band, emitting single photons at near-infrared (NIR) wavelengths (though visible wavelengths are present in different quantum dot systems). This system is operational only at cryogenic temperatures. Color centers, which are point defects in the lattice of a substrate, provide a different set of emission properties. They are often operable up to, and in some cases even beyond, room temperature, at a variety of wavelengths across the visible and NIR spectrum. For example, the negatively charged nitrogen-vacancy in diamond, usually called the NV center, occurs as a lattice defect in which two carbon atoms are replaced by a nitrogen atom and a vacancy [18]. The defect localizes electronic orbitals with energies within the diamond band gap, and electron transitions between these levels are tied to photon absorption and emission. Finally, quantum emitters usually have optically addressable spins, which can be exploited for quantum bit (qubit) and nanosensing applications.

The capability of a substrate to convert one color of light to another depends on the lattice symmetry. Frequency conversion is needed to connect parts of optical networks that operate at different wavelengths, as well as for light generation and detection at a new set of wavelengths [13]. Such frequency-mixing processes are nonlinear, and depend polynomially on the input power of light. Photonic applications often rely on low power consumption, which makes quadratic processes, such as the second harmonic generation utilized in green laser pointers, preferable in terms of efficiency. To facilitate these effects, the substrate has to exhibit second order nonlinearity, which is exhibited in non-centrosymmetric materials.

Table 1 summarizes the properties of commonly used photonic substrates.

|  | Si | GaAs | diamond | 6H/4H-SiC | 3C-SiC |
|---|---|---|---|---|---|
| Index | 3.5 | 3.4 | 2.4 | 2.6 | 2.6 |
| Band gap [eV] | 1.12 | 1.42 | 5.5 | 3.05/3.23 | 2.36 |
| Quantum emitter | - | Quantum dots | Color centers | Color centers | Color centers |
| Room T emission | - | No | Yes | Yes | Yes |
| Wafer-scale substrates | Yes | Yes | No | Yes | Yes |
| Sacrificial layer | SOI | AlGaAs | - | - | Si |

Table 1 – Properties of substrates commonly used in photonics.

*Silicon* is a ubiquitous substrate for many applications including electronics, microelectromechanical systems (MEMS) and photonics. Its availability in silicon-on-insulator (SOI) platform where a layer of silicon dioxide is sandwiched between a thin slab of silicon and a silicon wafer, and a variety of the industrially developed processing techniques, make silicon photonics possible [21]. This includes implementation of waveguides, microresonators, as well as optical modulators in many different geometries. Silicon's index of refraction provides high contrast to both air and glass. These devices are operable at NIR wavelengths due to the narrow band gap, and there is no second order optical nonlinearity. Silicon is usually assumed not to host quantum emitters, however, there have been recent studies of color centers in silicon operating at longer wavelengths (2.9 μm) and cryogenic temperatures featuring long spin coherence times [22]. The lack of light sources and detectors at these wavelengths, as well as their increased absorption in glass or native oxides, make the use of this platform technologically challenging.

*The III-V substrates* such as GaAs and GaP have high index of refraction and second order optical nonlinearity, they are hosts of quantum dots, and can incorporate sacrificial layers. InAs/GaAs quantum dot platform with AlGaAs sacrificial layer has been prominent in photonics research, for example for implementing suspended photonic crystal cavities with embedded quantum dots for cavity quantum electrodynamics [3] or interfacing telecommunication wavelengths with quantum dots operating at 900 nm [23]. GaP, on the other hand, has a wider band gap and provides opportunity for operation at visible wavelengths with low absorption [24]. III-V substrates are not considered CMOS-compatible due to the potentially unwanted doping of group-IV substrates, as well as due to the lack of high quality heteroepitaxial growth on silicon.

*Diamond*, which is considered an insulator due to its exceptionally wide band gap, has entered photonics as a host of visible and NIR color centers [25], though more recently it has been also studied in opto-mechanics as well [26]. Its color centers are active up to room temperatures, and can have spin coherence times of up to a second [18]. Methods for growing color center rich nanodiamond have been developed; centimeter scale single-crystal diamond substrates are also commercially available. Diamond is chemically inert and therefore nontoxic to cells, which has allowed for applications in biosensing [10]. The platform does not feature a sacrificial layer nor the second order optical nonlinearity.

*Silicon carbide* (SiC) has been present in power electronics and the MEMS industry for decades [27], but has entered photonics only in 2011, when its color centers [28] and second-order nonlinear optical processes [29] have been studied at the nanoscale. This material occurs in about 250 different polytypes [30], however, three of them have been prominent in industry and research: the bulk hexagonal polytypes 4H-SiC and 6H-SiC, and the cubic 3C-SiC. Silicon carbide combines the desired properties of all the previously mentioned photonics materials: wide band gap, room temperature active color centers, long coherence times, second-order optical nonlinearity, and favorable opto-mechanical properties. It is available in many forms, from nanoparticles to 6-inch wafers. Due to its symmetry, 3C-SiC can be heteroepitaxially grown on silicon [31], which provides a direct sacrificial layer. 4H- and 6H-SiC layers can be bonded to silicon dioxide through the so-called Smart Cut process [29].

Here, we will discuss color centers and nanofabrication techniques in silicon carbide and diamond.

## 3. Color Centers and Their Characterization

Color centers have gained prominence as solid state quantum emitters due to their exceptional brightness, room temperature operation, optical control of electron and nuclear spins and angstrom-scale dimensions. Their applications range from single photon sources to nanoscale sensors.

A color center is a point defect in a lattice of a semiconductor that has optically addressable electronic transitions [32]. They can be intrinsic, e.g. a vacancy or a misplaced atom, or extrinsic, when an atom of a different species is incorporated into the lattice. Point defects in a lattice break its translation symmetry which means the local electronic levels are not solutions to the Bloch equation and that electrons can be confined into localized orbitals. If ground and excited electronic state are both within the band gap of the host material, they represent an isolated two-level system and the electron can be optically addressed using sub-band gap frequencies. The emission spectrum consists of two features (Fig. 1a): the narrow lorentzian peak called zero-phonon line (ZPL), arising from the zero-phonon transition, and a phonon side-band, which consists of the transitions to other vibronic levels of the ground state. Emission branching into ZPL (Debye-Waller factor) increases at cryogenic temperatures. Color centers are emitters of single photons, both at cryogenic and at room temperatures. This makes them excellent candidates for integrated photonics applications.

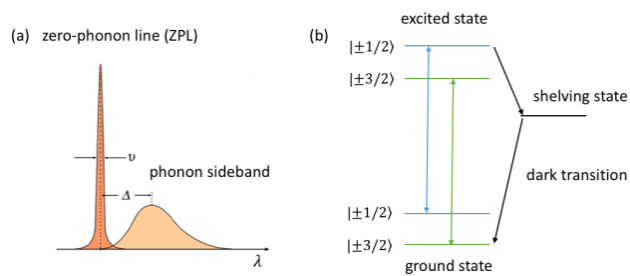

Figure 1 – (a) Elements of color center emission; figure adapted from [33]. (b) Energy levels of $V_{Si}$ in 4H-SiC in zero magnetic field.

Electron spin plays an important role in optical processes in color centers. Understanding spin-dependent electronic levels and their symmetry helps explain the emission peaks and their intensities in the experimental spectra. Here, we will use an example of the silicon vacancy ($V_{Si}$) color center in 4H-SiC [34] whose energy levels are illustrated in Figure 1b. Due to its electron spin of $S = 3/2$, the ground and excited level are split into ±1/2 and ±3/2 spin projections in zero magnetic field. Direct optical transitions between ground and excited state are spin-preserving. The intermediate shelving state that is non-radiatively coupled through spin-orbit interaction to one of the excited states, provides the mixing of the polarizations with preferential direction toward the ±3/2 projection, therefore enabling a polarization mechanism. The shelving state, also called dark state, allows an electron to change its spin in the de-excitation process while emitting a photon of lower energy which can be filtered out in the experiment. The spin state, also called dark state, that is more prone to the spin-flip exhibits lower intensity of photoluminescence, which provides a readout mechanism for the spin orientation. Applying microwave pulses at the frequency of the ground state splitting causes the spin to precess from one orientation to the other in a coherent fashion. These are the basic principles of initialization, readout and coherent control of electron spins in color centers used for spin qubits and magnetic sensing. In the case of qubit applications, the states |0> and |1> are represented by energy states ±3/2 and ±1/2, respectively.

Color centers can be naturally occurring, generated by incorporation of other atoms during growth of the substrate [35], or by electron [36] or neutron [37] irradiation of samples. Strain, electric field, and other non-uniformities in the substrates cause color centers in an ensemble to experience different environments which is reflected in a variation of their properties such as emission wavelength. Spin-coherence time is affected by the nuclear bath, which can be controlled through isotopic purification of the sample where only nuclear spin-zero isotopes are used during the epitaxial growth of the substrate.

To characterize the properties of the sample, we have used Photoluminescence and Optically Detected Magnetic Resonance measurements.

*The Photoluminescence* (PL) is a method where the absorption of a photon excites the system which then spontaneously luminesces. The frequency of the emission peaks provides the identification of a color center system. The excitation is performed with an above-band laser delivered to the sample through a confocal microscope configuration and filtered from the collected signal by dichroic mirrors and long-pass filters. Due to the random positioning of color centers in the substrate the 2D scanning capability is beneficial in characterization of the sample. This is realized through a rotating mirror followed by a 4f confocal setup and a high numerical aperture objective lens.

To characterize the emitter lifetime, the sample is pumped with the pulsed excitation and the signal is sent to the single photon counting module synchronized with the laser repetition rate. The measurement builds up the statistics of the photon detection in terms of the delay relative to the triggering signal, which corresponds to the excited state occupation probability.

*The Optically Detected Magnetic Resonance* (ODMR) measurement provides an optical readout of the spin resonance. Here, microwave (MW) excitation is used to rotate the electron-spin and PL is used to readout the state of the system [36]. For color centers that possess a spin-polarizing dark state the ODMR signal represents a resonance in the optical signal as a function of MW frequency. The resonance occurs when the microwave frequency corresponds to the energy splitting between the ground states with different spin projections. Once this frequency is determined, MW pulses of arbitrary length can be used to perform rotations on Bloch sphere needed for qubit applications.

## 4. Color Centers in Diamond and Silicon Carbide

The negatively charged nitrogen-vacancy (NV) center in diamond has, until recently, been the dominant color center system in photonics [25]. Its ZPL at 637nm contains 3-5% of the emission. Single-photon generation rates are in the tens of kHz and have been enhanced by the use of nanopillar structures to 168 kHz [38]. The NV center in diamond has been utilized for quantum cryptography [39], and integrated systems with silicon nitride waveguides have been demonstrated [40]. This color center exhibits $S = 1$ ODMR resonance which has been utilized to develop nanoscale magnetic sensor with unprecedented spatial resolution [41]. Temperature sensitivity of the ODMR has been utilized for *in vivo* thermometry [10]. Long spin coherence times of up to a second [18] are enabling long-lived spin-qubits. One of the challenges in working with the NV center is its spectral instability [42], which has motivated a search for novel color center systems [43].

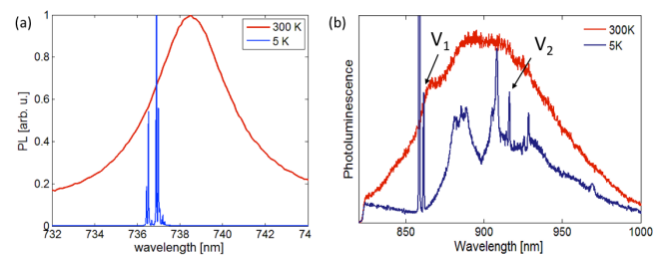

Figure 2 - Photoluminescence spectra. (a) Silicon-vacancy center in diamond exhibiting four ZPLs caused by spin-orbit coupling in each system. (b) Silicon vacancy center in 4H-SiC exhibiting two ZPLs from two non-equivalent positions in the lattice.

The negatively charged silicon vacancy ($SiV^-$) in diamond has emerged as an emitter with a large (70%) emission ratio into its four ZPLs at 738nm (Fig. 2a). Due to electron spin $S = 1/2$ the ground and excited state are split by 50 GHz and 250 GHz, respectively, by spin-orbit interaction. In addition, the ensemble linewidths are as narrow as 400 MHz at cryogenic temperatures [14], which qualifies these systems as nearly identical solid state quasi-atoms. An order of magnitude increase in the ZPL extraction and two to three orders of magnitude reduction in the ZPL linewidth compared to the NV center makes SiV a more attractive center for scalable systems. This is a result of the central symmetry of SiV leading to the absence of its permanent dipole moment and to the insensitivity to external electric fields. Additionally, reactive ion etching can be performed while maintaining the SiV center optical properties, which has been demonstrated with an array of nanopillars [35] and in nanobeam cavities [44]. These recent results pave a promising path for integrating nearly identical quantum emitters in the suspended photonic structures needed for cavity QED.

Chromium-related color centers have recently emerged as single-photon emitters at wavelengths around 760 nm [45]. Their structure has not been completely understood, however, there seem to be several types of ZPL transitions. These color centers have remarkable brightness of 500 kcounts/s.

Silicon carbide hosts numerous color center systems across its polytypes [46]. Emitters in 3C-SiC, such as the carbon-anti-site-vacancy ($V_CC_{Si}$) [47] and oxidation-induced centers [48] have exceptional brightness of around 1MHz count rates. The divacancy ($V_CV_{Si}$) in 4H-SiC gives rise to six ZPL transitions close to 1,100 nm wavelength, two of which are active and can be coherently controlled at room temperatures [28]. The silicon vacancy ($V_{Si}$) in 4H-SiC and 6H-SiC emits into 25-30 GHz narrow ZPLs capturing several percent of emission at cryogenic temperatures in their two and three lines [49], respectively, in the wavelength region 860-920 nm. These centers have been used as qubits [50] and magnetic field sensors [51]. In particular, we are interested in the 916nm ZPL (Fig. 2b) from $V_{Si}$ in 4H-SiC that has been demonstrated as a qubit at the level of a single color center [50].

## 5. Quantum Photonics in Silicon Carbide

Silicon carbide is an industrially mature substrate that has been used in power electronics [52] and microelectromechanical system (MEMS) development [27]. These technological developments have enabled a fast growth of silicon carbide color center photonics toward a scalable quantum platform. We utilize a commercially obtained 4-inch 4H-SiC wafer and develop electron beam lithography process to generate a matrix of nanopillars hosting individual silicon vacancy color centers [36, 53]. The nanopillar serves as a waveguide for the light emitted by its color center toward an objective. Compared to the previous approaches, this $V_{Si}$/4H-SiC platform provides advanced scalability, device footprint, optical interface and collection efficiency, while preserving optically induced electron-spin polarization properties. Therefore, the system can be applied as a scalable array of single photon emitters or spin-qubits.

The nanofabrication process is illustrated in Figure 3. A commercial 4H-SiC substrate was irradiated by 2 MeV electrons at a fluence of $10^{14}$ cm$^{-2}$ to generate $V_{Si}$ color centers sparsely. The substrate was covered in 300 nm aluminum to serve as a hard mask in the etching process. The electron-beam patterning of a negative-tone resist was used to define a matrix of nanopillars with diameters in the range 400 nm – 1 μm. Aluminum was etched in $Cl_2$ and $BCl_3$ plasma, and the pattern was transferred to silicon carbide in an $SF_6$ plasma process forming 800 nm tall nanopillars.

The pillar waveguiding of the color center emission is modeled using the Finite-Difference Time-Domain method which evolves Maxwell equations in the time domain (Fig. 3e). A continuous wave vertically-polarized dipole excitation is placed at the center of the pillar whose index of refraction is given by $n_{SiC}$ = 2.6, while the surrounding area holds $n_{air}$ = 1. The far-field image is analyzed to estimate the ratio of light that enters the NA = 0.65 numerical aperture of the objective. The theoretical increase in collection efficiency is up to two orders of magnitude, however, experimental results show only up to three-fold enhancement.

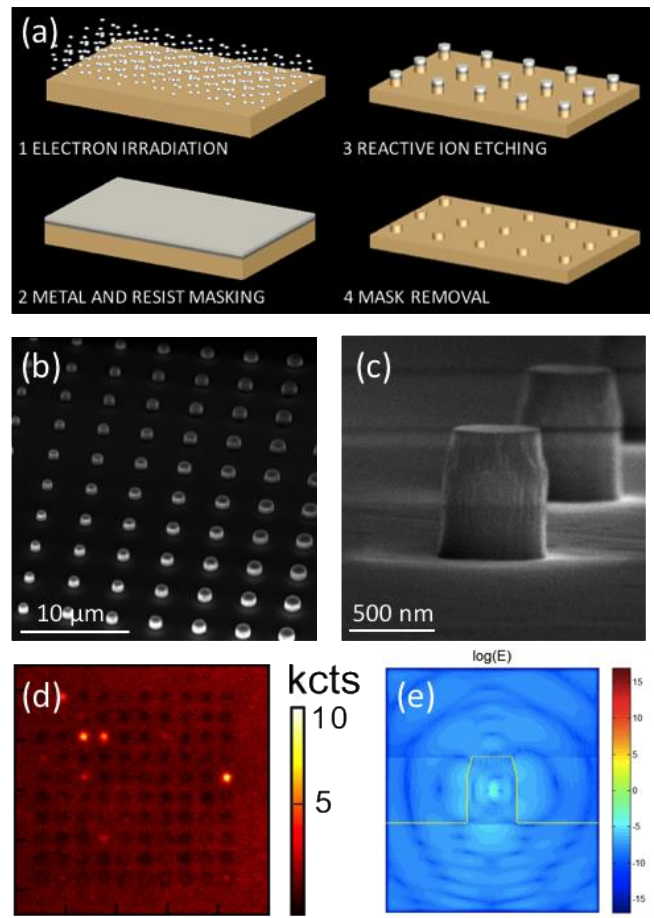

Figure 3 – (a) Lithographic process for fabrication of nanopillars in 4H-SiC. (b) SEM images of fabricated matrix of 800 nm tall nanopillars with variable diameters. (c) SEM image of a 600 nm diameter nanopillar. (d) Photoluminescence measurement on a matrix of pillars. (e) Finite-Difference Time-Domain model of electric field propagation from a dipole emitter placed in the center of a 600 nm wide nanopillar.

The photoluminescence map of a nanopillar matrix (Fig. 3d) shows that roughly 4% of the pillars contain a well-positioned color center, which is further confirmed as a single photon emission and ODMR spin resonance. This confirms the platform's capability to support scalable and efficient single-photon sources and quantum bits for applications in quantum technologies. The approach can be extended to incorporate other color centers in bulk 3C-, 4H- and 6H-SiC samples [47, 54, 55, 56, 57]. In addition, silicon carbide is a biocompatible substrate [58] and can be directly interfaced with cells for potential magnetic [37, 59] and temperature sensing applications [60], whose proof-of-principle demonstrations have recently been reported.

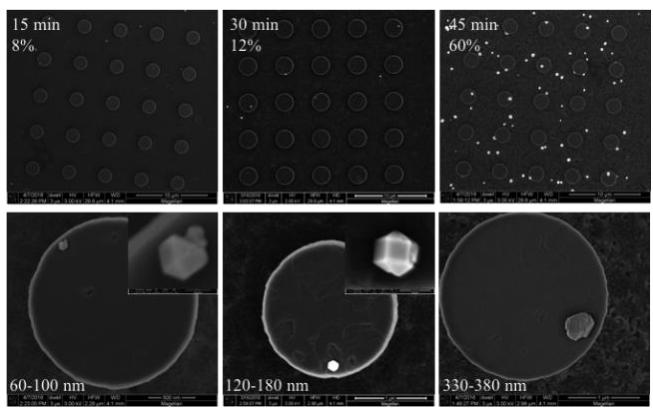

Figure 4 – Three generations of nanodiamond growth. The increasing CVD growth time results in larger nanodiamonds and a higher yield of hybrid devices.

## 5. Integration of Diamond Color Centers with Silicon Carbide Photonics

Diamond hosts a variety of color centers that are applied in quantum sensing and networks. Their integration with photonic devices is a crucial step in the development of efficient and scalable systems. One of the main challenges in diamond photonics is in the nanofabrication techniques. Since diamond does not grow heteroepitaxially on a sacrificial layer, suspended devices have to be developed through a complex set of fabrication steps, such as the Faraday-cage angled etching [61] or quasi-isotropic etching [62]. While these approaches yield noteworthy photonics results [26, 44, 63, 64], the device geometries that can be realized are limited to a subset of realizable shapes.

Hybrid platforms, where color center-hosting diamond is interfaced with a different substrate provide more flexibility in device geometry, as has been shown in gallium phosphide-diamond systems [65]. However, the index of refraction contrast between these two materials keeps the electromagnetic field localized to gallium phosphide and the color centers experience only an evanescent tail of the field. Additionally, III-V substrates might result in an undesirable doping of diamond.

We explore a hybrid photonic platform between silicon carbide and the color center-hosting nanodiamond. These two substrates have very similar optical properties with index contrast of only 0.2 which allows electromagnetic field to penetrate into both substrates, and wide band gaps which allow propagation of color center emission without absorption. Moreover, 3C-SiC is grown heteroepitaxially on silicon therefore providing a sacrificial layer for implementation of various suspended photonic geometries [66, 67]. Here, we show how hybrid microdisk resonators can facilitate resonant enhancement of diamond's silicon-vacancy and chromium-related color centers.

Using diamondoid-seeded chemical vapor deposition (CVD) [35, 68] nanodiamonds are grown on top of pre-fabricated 3C-SiC microdisk resonators on a silicon wafer. Silane is added to the process to facilitate the formation of SiV color centers, while the residual chromium in the chamber generates Cr-related center generation. Figure 4 shows the fabricated hybrid structures where the time of growth defines the nanodiamond size and the yield of the hybrid devices. The size of a nanodiamond correlates with the number of color centers it hosts, from several to hundreds. Finite-difference time-domain modeling reveals that larger nanodiamond more intensely perturbs the whispering gallery mode supported in the microdisk resonator, thus reducing its quality factor, but that they also result in higher penetration of the resonant electromagnetic field in the color center region of the hybrid device.

Figure 5 demonstrates the three-fold resonant enhancement of an individual SiV color center emission in an 80 nm diameter nanodiamond. The enhancement in 350 nm diameter nanodiamonds was measured to be as high as five-fold. This behavior was observed with Cr-based centers as well. The main origin of the enhancement is the Purcell effect that arises due to the small mode volume and high quality factor of the microcavity [68].

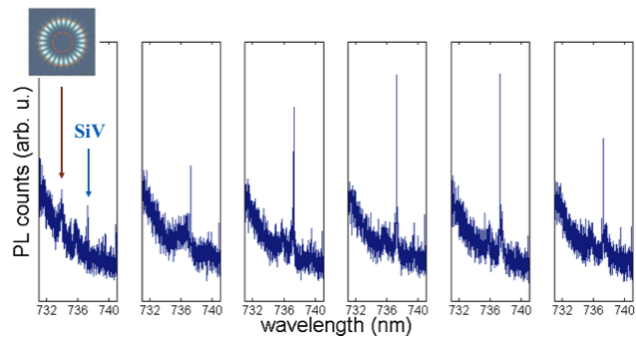

Figure 5 – Tuning of the microdisk whispering gallery mode over the silicon-vacancy center wavelength showing the resonant enhancement of the SiV emission.

This result showcases the potential of using a silicon-based SiC photonic platform to resonantly enhance diamond color center emission and, therefore, increase the speed of quantum channels. The approach can be readily extended to other device geometries and other diamond color centers.

## 6. Conclusion

Color centers in silicon carbide and diamond are promising solid state light emitters and spin-qubits with applications in quantum communications and sensing. Their integration with photonic devices is key to the development of arbitrarily complex quantum systems. Our results address technological progress in the incorporation of color centers into the waveguiding and resonant structures. The future of this line of research includes all-optical spin manipulation and an expansion of spin-photon interfaces to on-chip quantum simulators.